\documentclass[a4]{article}

\usepackage[lofdepth,lotdepth]{subfig}
\usepackage{graphicx}

\usepackage{algorithm}
\usepackage{algpseudocode}
\usepackage{color}
\usepackage{cite}
\usepackage{amssymb}
\usepackage{gensymb}
\usepackage{hyperref}
\usepackage{multicol}

\usepackage{color}
\usepackage{bm}
\usepackage[normalem]{ulem}

\newcommand{\Ohmmm}{$\Omega$m$^2$ }

\begin{document}

\title{{\bf Electro-thermal quench in metal-insulated nested REBCO coils for magnets over 40 T}}

\author{Anang~Dadhich$^1$, Philippe~Fazilleau$^2$ and	Enric~Pardo$^1$$^*$\\
\\
$^1$Institute of Electrical Engineering, Slovak Academy of Sciences,\\
Bratislava, Slovakia\\
$^2$Université Paris-Saclay, CEA, Département des Accélérateurs,\\
de la Cryogénie et du Magnétisme, 91191, Gif-sur-Yvette, France\\
$^*$Author to whom correspondence should be addressed\\
(enric.pardo@savba.sk)
}% <-this % stops a space

\maketitle

\begin{abstract}
Superconducting high field magnets have the capability to generate over 40 T, with multiple existing practical applications globally. However, at such high magnetic fields, these magnets are prone to rapid electrothermal quench which can affect the continuous operation of such magnets. A nested stack configuration, with multiple HTS inserts inside a LTS outsert, can be used for better thermal stability and compact design. We have performed detailed multiphysics quench analysis of such a nested stack high field magnet design under SuperEMFL project using our in-house software, which considers screening currents. Through various case studies, we have identified various weak spots in such a magnet, where thermal quench can be the most detrimental for magnet operation, and various ways are suggested to overcome this important issue.  
\end{abstract}

\section{Introduction}

Coils made of REBCO High Temperature Superconductors (HTS) can generate high magnetic fields thanks to their high engineering current density. Thus, {REBCO} coils have many {interesting} applications in high field magnets \cite{hahnS2019Nat, baiH2020IESa, kimK2020IESa, lecrevisseT2022SSTa, durochatM2024IES, superEMFL, awaji2S021IES, utoT2025IES, superEMFL, sharmabookMag, wangQ2021SST, wangK2022SSc}, medical research \cite{yanagisawaY2022SST, mansojimenoM2023NMR, parkinsonBJ2017SST, batesS2023MRM}, fusion \cite{mitchellN2021SST, haackJ2025ASE}, {aircraft electric propulsion} \cite{haranKS2017SST, pardoE2019IES, grilliF2020JCS, wengF2020SSTa, ybanezL2022JCS, kalsiSS2023IES, miyazakiH2024IES}, {space electric propulsion} \cite{bogelE2022IPEC, olatunjiJR2024IES}, and {other} rotating machines \cite{prajzendancP2025Ene, abrahamsenAB2010SST, songX2019IEC}. For high field magnets, pancake HTS coils are used, and these magnets are able to generate over 30-40 T magnetic field \cite{fazilleauP2024IES, durochatM2024IES, dadhichA2024SSTa}. For example, the Nougat project's HTS magnet (insert) was designed to generate 10 T, and inserted in a 20 T resistive magnet outsert, to give a combined 30 T field \cite{fazilleauP2020Cry}. Using metal insulated winding at 4.2 K, the Nougat magnet measured up to 32.5 T \cite{lecrevisseT2022SSTa}. For SuperEMFL project, which intends to use a Low Temperature Superconductor (LTS) outsert instead of resistive magnet with an HTS insert, we aim to generate 32 T and 40 T, through different designs \cite{superEMFL, durochatM2024IES, fazilleauP2024IES, pardoE2024SSTa, dadhichA2024SSTa}. Globally, other high field magnets are also able to generate fields in this range, or are currently under development, such as the LBC hybrid demonstrator magnet (45.5 T) in Florida, USA \cite{denoudenA2016IES, hahnS2019Nat}, CHFML hybrid magnet (45 T) in Hefei, China \cite{wangQ2021SST}, HFLSM magnet (25-30 T) in Tohoku, Japan \cite{awaji2S021IES, takahashiK2024IES}, Grenoble hybrid LTS magnet in France (43 T) \cite{pugnatP2022IES}, {among other}.

Along with the aforementioned benefits, superconducting high field magnets also present many challenges. Firstly, superconductors are diamagnetic materials, which present magnetization and screening currents. These generate high AC losses during the ramping up of the magnet \cite{pardoE2016SST, pardoE2024SSTa, dadhichA2024SSTa}. Furthermore, when the current in the magnet goes over the critical current (${I_c}$) of the HTS, resistive heat is generated, which can lead to thermal runaway or 'quench', that risks the burning of the material if not controlled. Localized damage or inhomogeneity of material can also cause thermal hotspots, leading to quench \cite{liY2014JCS, badelA2019SST, chenJ2023SST, gavrilinAV2024IES, pardoE2026RIE}. Therefore, a high field magnet design needs to be thermally sound, with fail-safe techniques like voltage limiting, which can work for high field magnets using non-insulated or metal insulated windings \cite{pardoE2026RIE}. Adding an insert to high field outsert magnet tackles many of these issues, such as better thermal management (for instance, LTS outsert and HTS insert can have their own cooling media and cryogenic systems), scalability and ease of repair (coils can be replaced much easily in the case of damage and upgrade), improving the mechanical strength, and developing a compact design to generate high magnetic fields at the center of magnet as compared to single pancake stacks or solenoids \cite{dongF2025SST, zhangX2023IES, shaoL2023IEA}. This design can be further improved by inserting multiple HTS stacks ('or nested stacks') in LTS outsert to give better thermal stability (different inner stacks quench at different times, and can be thermally managed better) \cite{durochatM2024IES} and mechanical strength (as the multiple pancake coils can share the less turns, i.e. less stress accumulation). A multiple nested inner stack system, with 3 nested HTS coils, was tested recently at MIT for NMR magnets, and serves as a good demonstrator for this technology \cite{michaelPC2019IES}.

Also, as we see that the design of high field magnets involves complex and coupled multiphysics, a strong and fast software is required in the superconducting community to analyze these cases, as the commercial software can be very limited and slow in simulating these issues \cite{dadhichA2024SSTa}. This is mainly because of the high number of pancake coils in the system (at least 15), and each consisting of 200 or more turns. Furthermore, most electrothermal quench models consider uniform current density in HTS tapes due to ease of modeling and fast computation times \cite{stenvallA2023book, botturaL1996JCP, badelA2019SST, dongF2022APL, vitranoA2023IES}, but that misses out on additional AC losses generated by screening currents, which contribute significantly to the temperature rise in such high field magnets \cite{pardoE2026RIE, dadhichA2024SSTa}.

A nested stack system is currently being designed under SuperEMFL project, which involves 2 HTS nested stack insert inside a LTS stack outsert, and can generate higher fields at center (up to 45 T) as compared to normal solenoid designs \cite{durochatM2024IES}. This paper is focused on detailed multiphysics analysis (electrothermal and electromagnetic) of this nested stack system, using our in-house software that considers screening currents. Under this analysis, we have focused on thermal quench of the HTS inserts under different damage circumstances. Voltage limiting features are also applied to the design in the event of quench, to control the thermal runaway and letting the magnet to operate at lower power output. Doing so, we have identified various thermally weak points in such high field nested stack systems from where electrothermal quench can be detrimental to magnet operation.

\begin{figure}[tbp]
	\centering

{\includegraphics[trim=0 0 0 0,clip,width=12 cm]{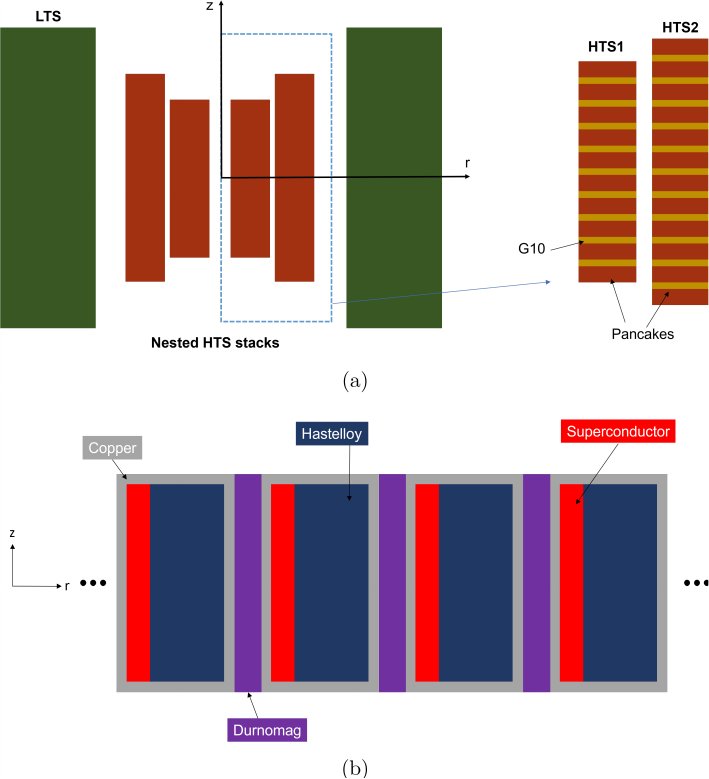}}

\caption{(a) Full scale magnet and its cross section, which shows nested HTS1 and HTS2 pancake coil insert stacks inside LTS outsert magnet. (b) Cross section of turns in pancakes, including the isolating Durnomag layer between 2 tapes. The sketches are only for representation and not to scale.  }
\label{crossSec}
\end{figure} 

\section{Modeling Method}

An in-house 2D axi-symmetric software is used to calculate different cases for this paper. The electromagnetic calculations are performed by the Minimum Electro-Magnetic Entropy Production (MEMEP) method, which is based on variational principles \cite{pardoE2015SST, pardoE2017JCP, pardoE2024SSTa}. {This method takes both screening currents and radial currents into account.} The thermal calculations are performed using explicit Finite Difference (FD) method. MEMEP and FD are coupled together to give accurate results (MEMEP-FD). This method considers temperature dependent electromagnetic and thermal properties, and this coupling is shown in our previous works \cite{dadhichA2024SSTa}. These methods {have been} also benchmarked for pancake and racetrack coils with various mainstream methods and commercial software {based on Finite Element Method} to prove its validity \cite{pardoE2023IES, dadhichA2024SSTb}. Additionally, voltage limiting features are added as a fail-safe method to limit the current input in the case of quench \cite{pardoE2026RIE}. Each simulation for this paper took around 3 days to complete in a standard desktop computer, which is fast considering the high number of pancakes and turns per pancake in the nested stack system (parameters described in Table \ref{param1}).

\section{Modeling parameters}

\begin{table}
\centering
\caption{{Design parameters of the HTS metal-insulated nested stack} \cite{durochatM2024IES}. {Copper thickness refers to the total thickness across the tape.}}
\label{param1}
\footnotesize
\begin{tabular}{@{}llll}
\hline
\hline
Nested stack dimension 	& Values				&	Material dimensions  &   Values  \\
\hline
Internal Diameter				& 50 mm         & REBCO           &  2 $\mu$m    \\
External Diameter				& 216.2 mm      & Substrate (Hastelloy)     &  54 $\mu$m   \\
HTS1 pancakes						& 42            & Copper        &  10 $\mu$m   \\
HTS2 pancakes						& 44            & Durnomag      &  30 $\mu$m    \\
HTS1 turns {per pancake}	& 360           & G10 width     &  0.5 mm   \\
HTS2 turns {per pancake} 	& 260           & Pancake/tape width &  6 mm     \\ 
LTS magnetic field   		& 15 T         & Durnomag resistivity     & 10$^{-6}$ $\Omega\cdot$m$^2$           \\ 
Limiting voltage     & 2.5 V         &  Nominal current  & 231.2 A           \\    
\hline
\hline
\end{tabular}\\
\end{table}
\normalsize

{We consider a user magnet design from the SuperEMFL project (basic design N1 for the 40 T magnet)} \cite{durochatM2024IES}. The LTS outsert is a 15 T/250 mm bore magnet from Oxford Instruments \cite{fazilleauP2024IES}. These summarized parameters are shown in Table \ref{param1}. The complete magnet system consists of LTS outsert, and {an HTS insert made of two nested stacks of metal-insulated pancake coils} with their own power (electrical) circuit, as shown in Figure \ref{crossSec}(a). The HTS1 stack is connected to the HTS2 stack in series, and the LTS stack provides a background field of around 15 T to the nested HTS stacks. This whole system is initialized at 4.2 K, i.e. liquid Helium temperature. 

For the calculations, Theva APC tape is considered {(APC stands for Advanced Pinning Center)}, and its dimensions are given in Table \ref{param1}. The cross section in the thickness of the tape is shown in Figure \ref{crossSec} (b). The temperature and magnetic field dependent electromagnetic and thermal properties of the tape layers, including the tape design and assumptions can be seen in our previous works \cite{dadhichA2024SSTa, pardoE2026RIE}. {Neighboring} turns are separated by a metal layer of Durnomag, which allows for radial currents between layers. A complete $J_c(B,T,\theta)$ dependence of Theva tape is considered for calculations \cite{senatoreC2024SST, pardoE2026RIE}, which allows to consider the decrease in the localized current density with magnetic field and temperature.

{We analyze quench due to a group of damaged turns.} {As a worst-case scenario, we consider that $J_c$ drops to 10 \% of its original value in the damaged turns.} The damaged group of turns is 41 to 50 turns (10 turns in total), and the specific pancake where the damage occurs is specified in the next sections, case wise. The reason for using 10 turns as a group, is due to the homogenization of 10 turns as one homogenized block in our simulations to improve the speed of the calculations \cite{pardoE2024SSTa}. The magnet is set to be running at a steady nominal current (231.2 A) as an initial condition, when damage appears. Hence, the quench is analyzed from this point in time, with a fixed time step of 10 milliseconds. These studies also assume adiabatic conditions at the boundary of the nested stacks, i.e. no heat is exchanged with the external coolant or extracted from the system. Lastly, the turn-to-turn contact resistivity of the winding (or Durnomag layer here) is considered to be $10^{-6}$ {\Ohmmm} (typical of metal insulated windings \cite{genotThesis2021}) in all cases, unless specified.  

Additionally, HTS1 and HTS2 are thermally and electrically isolated from each other, and {hence} they interact only through inductive effects. The current density and temperature profiles shown in the next section are for the cross section as shown in Figure \ref{crossSec} (a).

\section{Results and Discussion}

The electrothermal quench in the HTS nested stacks is studied in the next sections, for different case scenarios and parameters. Firstly, we study the effect of the location of the damage (weak spot) in the inner HTS stack. All these studies are performed considering the screening currents distribution in the simulations{. This a key feature for quench analysis, since assuming uniform current density under-estimates electrothermal quench propagation speed \cite{pardoE2026RIE}.} Thus, a case with uniform current density ($J$) has been added in the 1st study to show {the impact of screening currents} for nested stacks. Secondly, {we make a parametric study of the contact resistance between turns}. Lastly, we {analyze} the effect of damage at the outer stack (HTS2).

\subsection{Study 1: Weak spot {location}}

\begin{figure}[tbp]
	\centering

{\includegraphics[trim=0 0 0 0,clip,width=10 cm]{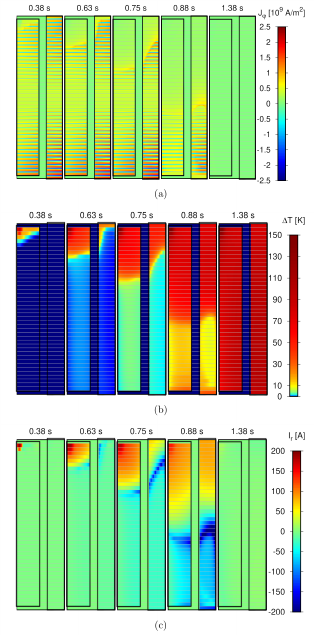}}

\caption{Hotspot analysis when damage is at the top pancake at HTS1. Figure shows (a) current density profiles, (b) change in temperature, and (c) increase in radial currents up to 1.38 s.}
\label{weakSpot_top}
\end{figure}

\begin{figure}[tbp]
	\centering
	{\includegraphics[trim=0 0 0 0,clip,width=10 cm]{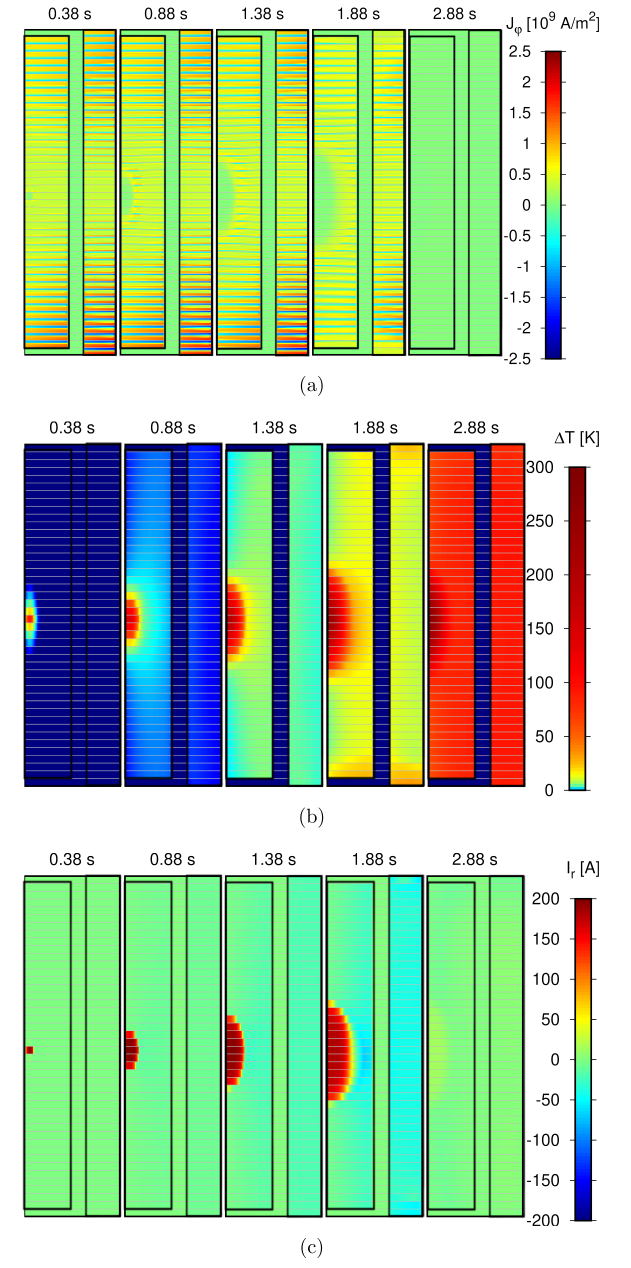}}

\caption{Hotspot analysis when damage is at the middle pancake at HTS1. Figure shows (a) current density profiles, (b) change in temperature, and (c) increase in radial currents up to 2.88 s.}
\label{weakSpot_mid}
\end{figure}

\begin{figure}[tbp]
	\centering

{\includegraphics[trim=0 0 0 0,clip,width=10 cm]{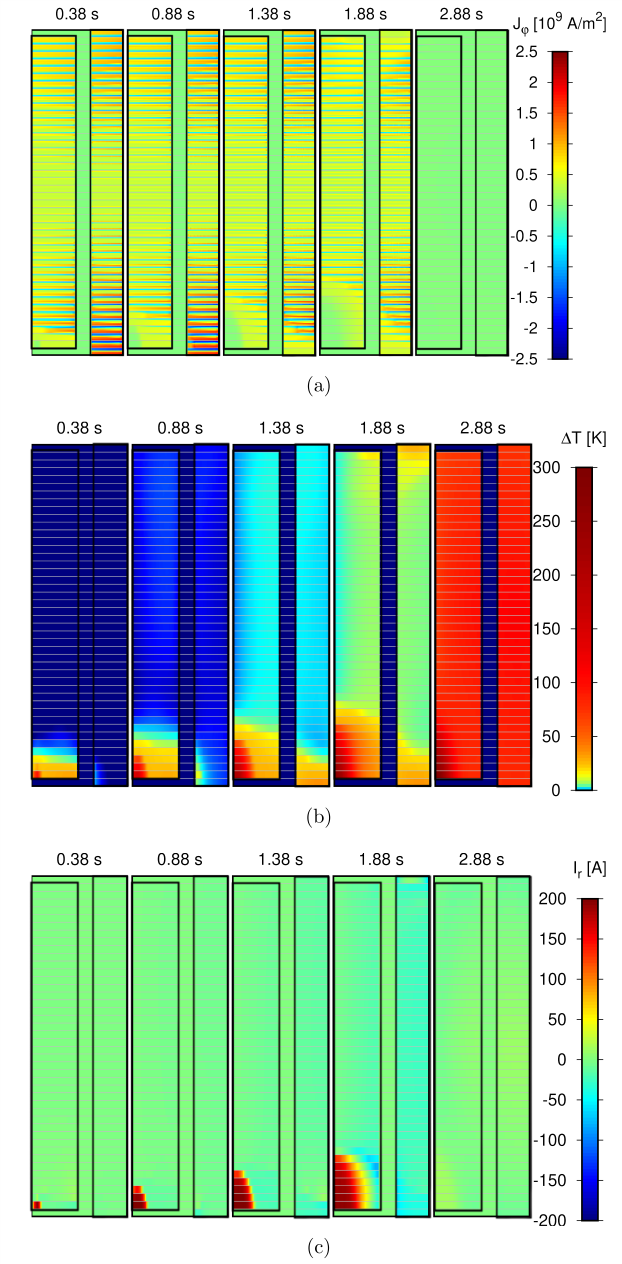}}

\caption{Hotspot analysis when damage is at the bottom pancake at HTS1. Figure shows (a) current density profiles, (b) change in temperature, and (c) increase in radial currents up to 2.88 s.}
\label{weakSpot_bottom}
\end{figure}

\begin{figure*} [tbp]

{\includegraphics[trim=0 0 0 0,clip,width=\textwidth]{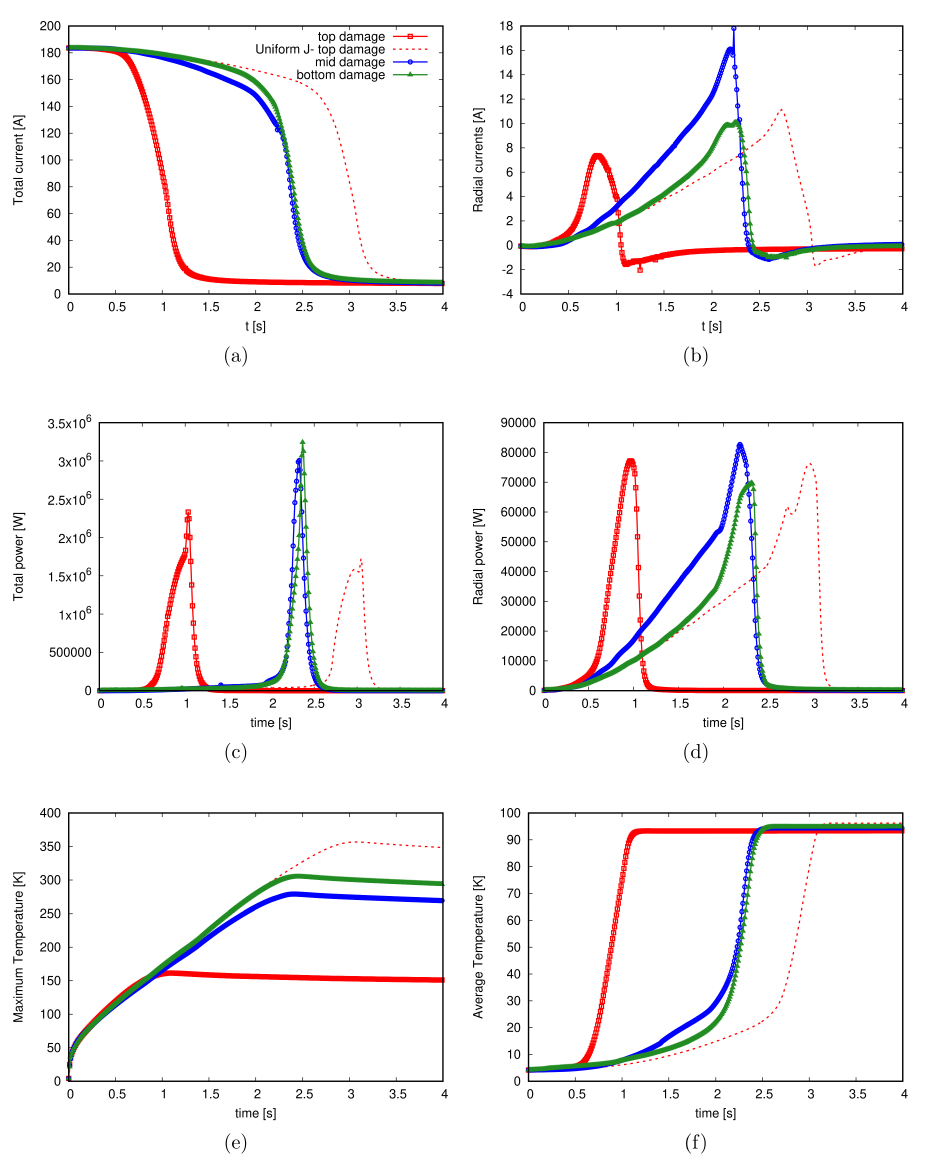}}

\caption{Nested stack behavior when damage occurs at different parts of the stack. Figures show  (a) Total current decay, (b)Average radial current in section, (c) Total power, (d) Power loss due to radial current, (e) Maximum temperature, and (f) Average temperature rise in the nested stacks after the onset of damage. It is seen that maximum temperature rise occurs when the damage is at the bottom of the stack, and hence it is considered as the worst case scenario (or our reference case). All figures use same legend as (a).}
\label{weakSpot_curves}
\end{figure*}

{This section analyses the impact of the weak spot location in the nested stack configuration}. For this purpose, damage is induced in 10 aforementioned turns in different pancakes in the inner stack (HTS1): top, bottom, and middle pancake. The results for this study is shown in figures \ref{weakSpot_top}-\ref{weakSpot_curves}.

Figures \ref{weakSpot_top}-\ref{weakSpot_bottom} show {azimuthal} current density, change in temperature ({$\Delta T=T-T_{\rm initial}$}, where {$T_{\rm initial}=4.2$} K), and radial current profiles when the damaged turns are at the top, middle, and bottom pancakes of the HTS1 stack, respectively. {At each turn, the total current is the azimuthal current plus the radial current. Then, the coil total current consists on the average azimuthal current plus the average radial current in the coil cross-section.} The following general behavior for these stacks can be seen from these profiles: the change in temperature is the highest in the damaged turns, which decreases the {azimuthal} current density in these turns and gives rise to the radial currents. This temperature rise propagates through the whole HTS1 stack, decreasing the critical current density everywhere (or transiting the superconductor to normal state ($T>T_c$)). As a consequence, radial currents rise. These radial currents further increase the power losses and temperature of the magnet, as we have seen in our previous work \cite{dadhichA2024SSTa}. Thus, quench occurs in the whole stack quickly due to this quench propagation. Another major reason for such high power losses is the presence of screening currents in the system, which gives rise to high AC losses \cite{pardoE2026RIE}. The swift changes in magnetic field from the HTS1 stack also affect the HTS2 inductively that contributes to its quenching, and it can be seen from these figures that both nested stacks quench almost simultaneously. Consequently, decrease in angular currents in HTS2 also contributes in inducing quench in HTS1.

These figures also show that the configuration where the damage is at the top of the HTS1 stack quenches much faster than the other 2 cases, which is around one second (see Figure \ref{weakSpot_top} (a)-(c)). Comparatively, the case with the damage at bottom, takes the longest time to quench (up to 3 seconds) as seen in Figure \ref{weakSpot_bottom} (a)-(c). However, the temperature rise is the highest in the latter case (up to 300 K, as also seen in Figure \ref{weakSpot_curves} (e)), as compared to the others. Then we can consider this as the worst case scenario, due to the higher chances that the HTS tape can experience damage by electro-thermal mechanical stress. Thus, the next studies in this paper considers damage at the bottom of HTS1 as the reference case for detailed calculations. 

The reason for the faster quench in the bottom damaged case is the $J_c(B,T,\theta)$ behavior of the considered Theva HTS tape. The angular dependence of $J_c$ is non symetric with respect to the tape plane, with angular deviation around 30°, and varies highly around the 90 degree field angles \cite{pardoE2026RIE}, which makes its $J_c$ much lower at the top of the nested stack as compared to the bottom. Due to this lower $J_c$ at the top, the nested stack quenches much faster at the top, while the quench also propagates faster. At the bottom, $J_c$ is higher, which keeps most of the pancakes superconducting. However, then the quench propagation is also slower which keeps the heat accumulated in the smaller damaged sections and it gives rise to higher maximum temperature. Other REBCO tapes, like Fujikura, do not have this asymmetry, and quench may propagate more symmetrically from both top and bottom \cite{pardoE2026RIE, dadhichA2024SSTa}.

Figure \ref{weakSpot_curves} shows several electrical and thermal quantities of the complete nested stack. Firstly, Fig. \ref{weakSpot_curves} (a) {shows} a drop in the total current due to voltage limitation and the appearance of non-superconducting parts due to electrothermal quench {(or the decrease of the $I_c$ of several turns below the operating current)}. Without voltage limitation, the total power would increase monotonically, leading to thermal runaway. There is rise in total power due to the current density {overcoming} local $J_c${,} which gives rise to temperature (Figures \ref{weakSpot_curves} (c) and (e)). This increase in temperature further reduces $J_c$, and hence there are more coil turns with current higher than local $I_c$, contributing further to the total power rise (mainly azimuthal). Also, this fast change in screening currents (azimuthal direction) causes significant AC losses \cite{pardoE2026RIE}. The azimuthal current {at each turn} gets converted to radial currents {during electrothermal quench; which} gives rise to the radial power{, or power due to radial currents} (Figure \ref{weakSpot_curves} (d)){. This radial power} also contributes to the temperature rise in {the} magnet \cite{pardoE2024SSTa, dadhichA2024SSTa}. Thse negative {average} radial currents seen in Figure \ref{weakSpot_curves} (b) are due to the induced effects due to the decrease in magnetic flux caused by the decrease in average angular current. Figures \ref{weakSpot_curves} (e,f) show that when damage appears at the bottom, electrothermal quench is the slowest but the maximum temperature is the highest. For all configurations, the average temperature overcomes the critical temperature, 92 K (Figure \ref{weakSpot_curves} (f)).

It is interesting to note that the case with damage at the middle of HTS1 stack shows the highest radial power and {radial} currents, and the maximum temperature for this case is almost the same as the reference case (Figure \ref{weakSpot_curves}). {The reasons are the following. First, the maximum temperature occurs at the damaged turns. Since their $I_c$ is well below the operating current, their loss is dominated by the turn resistance in the angular and radial directions and the instantaneous current. As electro-thermal quench propagates at an intermediate speed between the cases of damage on top and bottom, the coil current drops at an intermediate time. Therefore, the accumulated heat in the damaged turns is in between that for damage at top and bottom. Thus, the maximum temperature for damage in the the middle is in between. The highest radial power is due to the fact that, once the pancake with damage fully quenched, quench propagates both upwards and downwards, as well as radially.}   

We also show one case in figures \ref{weakSpot_curves} (a)-(f) considering uniform current density instead of screening currents in the simulation, where damage is at the top pancake in HTS1 (red dashed curve). Here, it is seen clearly that the quench predicted by such assumption is not accurate, as widespread quench appears after almost triple the time than when screening currents are present (red solid curve). It also shows the highest maximum temperature, which shows that this prediction may not be accurate, compared to when screening currents are present, which is more realistic and physical case. {This is consistent with our previous work in \cite{pardoE2026RIE}.}

\subsection{Study 2: Effect of contact resistance between turns}

\begin{figure}[tbp]
	\centering

{\includegraphics[trim=0 0 0 0,clip,width=10 cm]{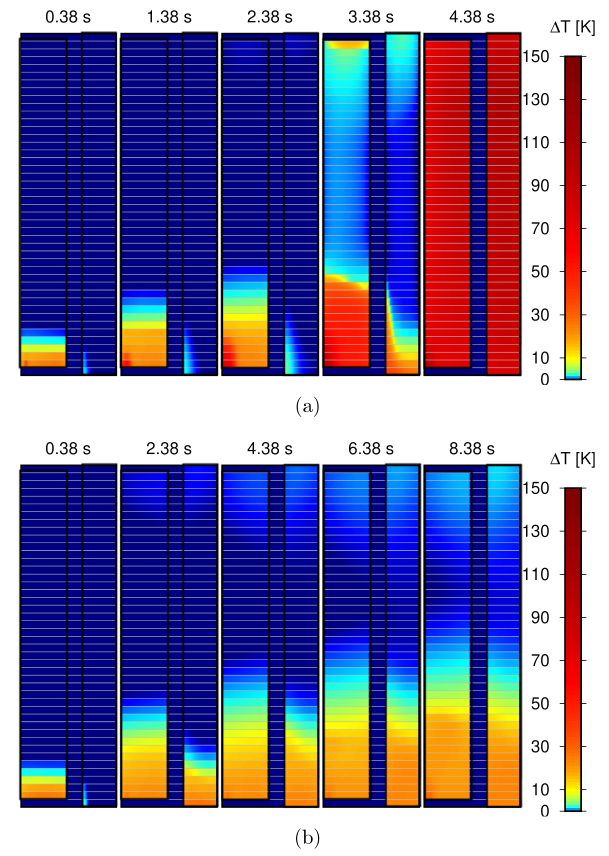}}	

\caption{Resistivity analysis with different turn to turn resistance. Figure shows change in temperature when turn-to-turn resistivity is (a) 10$^{-7}$ $\Omega\cdot$m$^2$, and (b) 10$^{-8}$ $\Omega\cdot$m$^2$  }
\label{resistance_dT}
\end{figure} 

\begin{figure*} [tbp]

{\includegraphics[trim=0 0 0 0,clip,width=\textwidth]{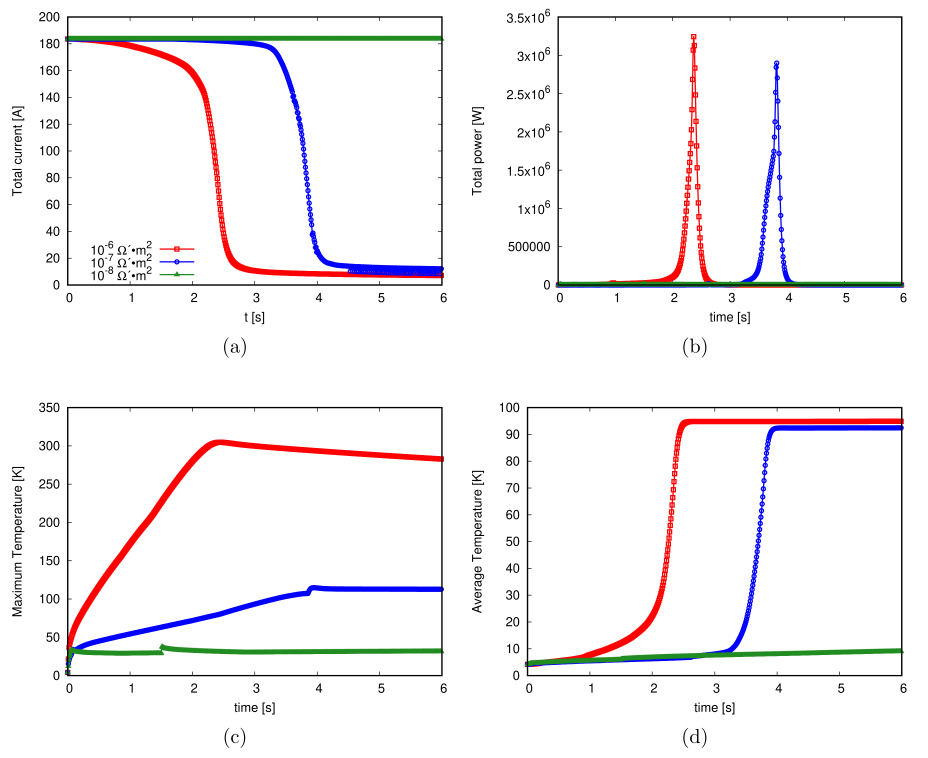}}
 
\caption{Resistivity analysis with different turn to turn resistance. Figures show (a) Total current decay, (b) Total power, (c) Maximum temperature, (d) Average temperature rise in nested stacks after the onset of damage. All figures use same legend as (a). }
\label{resistance_curves}
\end{figure*}

Since we observed in the previous section that the quench at the bottom of HTS1 stack is potentially the most damaging, next we study the effects on such quench when varying the contact resistance between turns. The previous study was done with turn-to-turn resistivity as 10$^{-6}$ $\Omega\cdot$m$^2$, and in the following we also consider 10$^{-7}$ $\Omega\cdot$m$^2$ and 10$^{-8}$ $\Omega\cdot$m$^2$.  The results for this study is shown in figures \ref{resistance_dT} and \ref{resistance_curves}.

Firstly, it is seen from Figures \ref{resistance_dT} and \ref{weakSpot_bottom}(b), that the case with {the} highest turn-to-turn {contact resistance} (10$^{-6}$ $\Omega\cdot$m$^2$) is the fastest to quench (in less than 3 seconds). The case with {contact resistance} of 10$^{-7}$ $\Omega\cdot$m$^2$ quenches in around 4.3 seconds, and the case with 10$^{-8}$ $\Omega\cdot$m$^2$ does not quench at all, even after 8 seconds (almost 4 times longer than the reference case). {However, the maximum temperature increases with the contact resistance.} We have seen {these effects} in our previous works {about single-stack inserts} \cite{pardoE2024SSTa, dadhichA2024SSTa}. The quench in nested stacks can be further reduced by using soldered coils or joints (with resistivities equal or below 10$^{-9}$ $\Omega\cdot$m$^2$), as being considered for certain REBCO magnets \cite{kosseJ2025SST, tsuiY2016SST}. {These may be complicated to build, but they are effective in avoiding delamination due to thermal stress during quench; thanks to lower thermal gradients \cite{kosseJ2025SST, munJ2020IES, tsuiY2016SST}.}
Soldered coils are better regarding thermal stabilization. However, they strongly limit the ramp rate, and may not be suitable for {the user magnets for the SuperEMFL project, which require relatively} fast ramp {rates}. 

Figure \ref{resistance_curves} shows further analysis for this study. It can be seen here that the non-insulated magnet shows almost no drop in the total current, or no rise in the total power, when compared to the other two metal insulated cases. The average and maximum temperature rises very slightly for the non-insulated case (Figure \ref{resistance_curves} (d) and (c)), whereas the metal-insulated magnet has already quenched within 4 seconds for both 10$^{-6}$ and 10$^{-7}$ $\Omega\cdot$m$^2$.

\subsection{Study 3: Damage at outer stack }

\begin{figure}[tbp]
	\centering

{\includegraphics[trim=0 0 0 0,clip,width=10 cm]{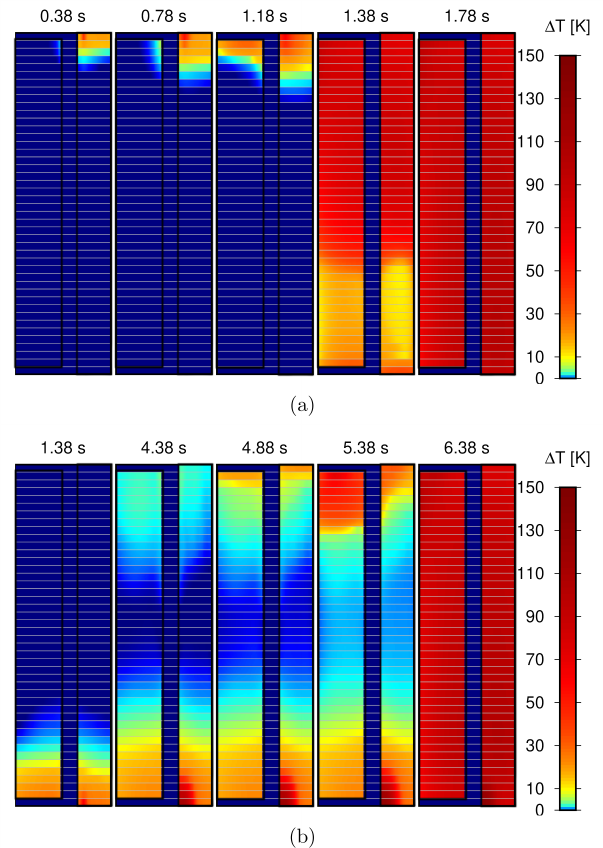}}
	
\caption{Nested stack behavior when damage occurs at different parts of the outer stack. Figure shows change in temperature when damage is at (a) top, and (b) bottom of HTS2 stack.}
\label{outer_profiles}
\end{figure} 

\begin{figure*} [tbp]

{\includegraphics[trim=0 0 0 0,clip,width=\textwidth]{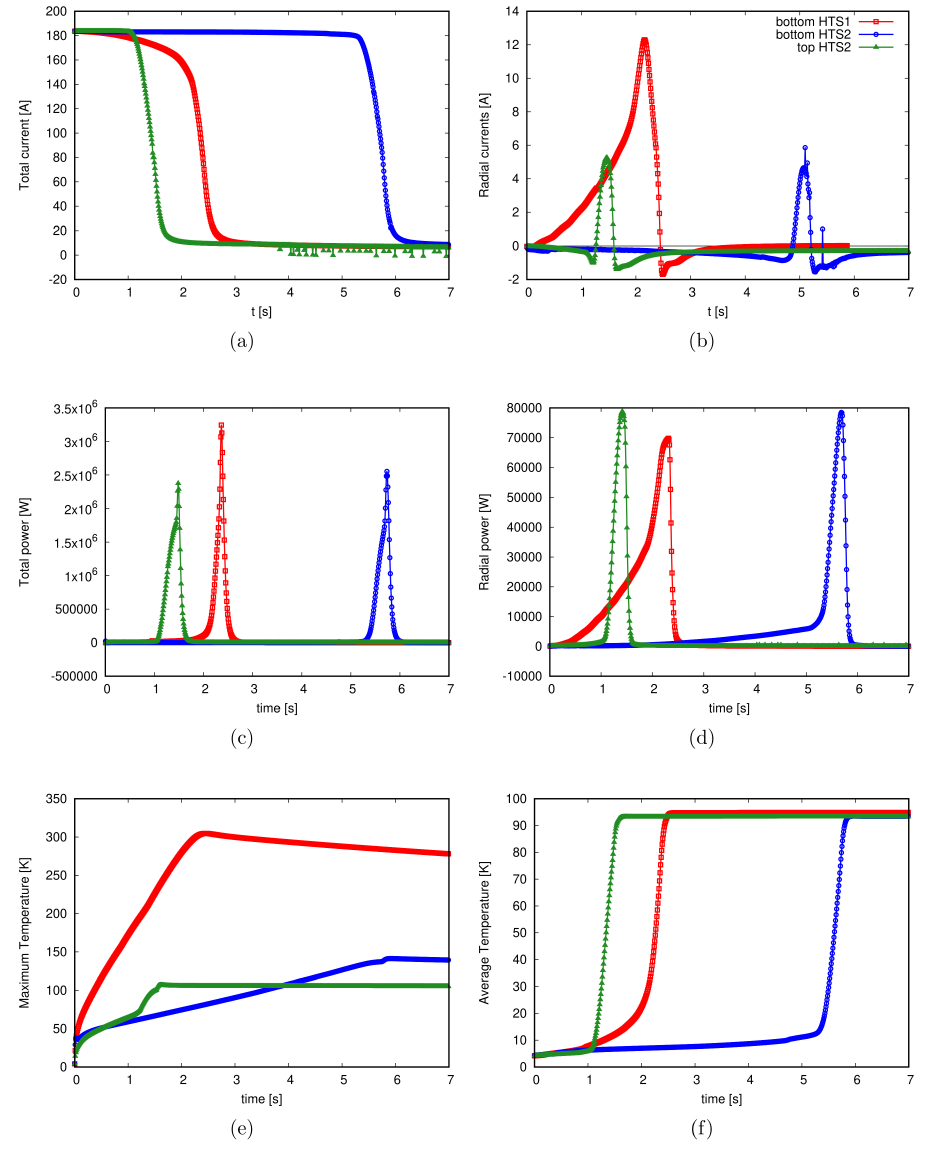}}	  
 
\caption{Nested stack behavior when damage occurs at different parts of the outer stack. Figures show  (a) Total current decay, (b) Radial current rise, (c) Total power, (d) Radial power, (e) Maximum temperature, and (f) Average temperature rise in nested stack after the onset of damage. It is seen that maximum temperature rise occurs when the damage is at the bottom of the HTS1 stack, which is the weak point of the whole system. All figures use same legend as (b). }
\label{outer_curves}
\end{figure*}

Lastly, we also check the effect on quench when it occurs due to damaged turns in the outer stack (HTS2). The results for this {configuration} is shown in figures \ref{outer_profiles} and \ref{outer_curves}. These results are compared with the reference case (damage at the bottom of HTS1).

The behavior is similar to when quench appears in the inner stack. The case with the damage at the top quenches faster (Figure \ref{outer_profiles} (a): 1.78 s) than the damage at the bottom of HTS2 stack (Figure \ref{outer_profiles} (b): 6.38 s). This consequently quenches the inner HTS1 stack as well, simultaneously. This is further verified from Figure \ref{outer_curves} (f), where the average temperature of the whole insert rises above critical temperature much later for the bottom damaged HTS2 stack than any of the other cases. The reason why quench for damage at the bottom of HTS2 appears later than if quench is at HTS1 is because the magnetic field at HTS2 is lower, and hence $J_c$ at HTS2 is higher. Therefore, the safety margin in the outer stack is higher, which causes the quench to appear later. 

The maximum temperature for both cases of quench in the outer stack is quite similar, with only around 20 K difference (Figure \ref{outer_curves} (e)), in contrast to the quench in the inner stack. The lower increase in temperature compared to the quench in the inner stack is because the radius of the damaged turns in the outer stack is higher. Then, the heat inertia is higher, and hence it requires more heat to increase the temperature. {In addition, the turn-to-turn resistance decreases with the radius for the same contact resistance.} Also, similar radial current and radial power rise for these 2 cases (Figure \ref{outer_curves} (b) and (d)) contributes to the temperature rise. The reference case still shows the highest maximum temperature of around 300 K, and thus it is the most sensitive point of such a magnet design, which can quench the whole magnet with very high temperatures.

\section{Conclusion}

A nested stack configuration with two HTS REBCO nested stacks insert {within} a LTS outsert is succesfully analyzed numerically for multiphysics quench, using our in-house software. It is seen that the bottom pancake of {the} innermost stack (HTS1) is the most sensitive to the electrothermal quench, showing the highest maximum temperature in all cases. Thus, precautionary measures should be taken to make this section thermally stable. This effect can be further reduced using lower resistivities between turns (below 10$^{-7}$ $\Omega\cdot$m$^2$) for successful long term operation of such high field magnets. However, lower contact resistance between turns will limit the ramp speed of high-field magnet, and hence an optimum value should be obtained for the particular use of the magnet.

\section{Acknowledgements}

We acknowledge Oxford Instruments for providing details on the cross-section of the LTS outsert. This {work} has received funding from the European Union's Horizon 2020 research and innovation programme under grant agreement No 951714 (superEMFL), and the Slovak Republic from projects APVV-24-0654 and VEGA 2/0098/24. {Research partially funded} by the EU NextGenerationEU through the Recovery and Resilience Plan for Slovakia under the project No. 09I04-03-V02-00039. {Part of the research results were obtained using the computational resources procured in the national project National competence centre for high-performance computing (project code: 311070AKF2) funded by the European Regional Development Fund, EU Structural Funds Informatization of society, Operational Program Integrated Infrastructure.} Any dissemination of results reflects only the authors' view and the European Commission is not responsible for any use that may be made of the information it contains. AD acknowledges the Schwarz Fund from the Slovak Academy of Sciences.

\end{document}